\documentclass[preprintnumbers,twocolumn,prx,amsmath,amssymb,longbibliography,10pt,aps]{revtex4-2} 

\usepackage{graphicx}
\usepackage{latexsym}
\usepackage{amsmath}
\usepackage{wasysym}
\usepackage{amsthm}
\usepackage{amsbsy}
\usepackage{amssymb}
\usepackage{epstopdf}
\usepackage{enumerate}
\usepackage{setspace}
\usepackage{dcolumn}
\usepackage{bm}
\usepackage{slashed}
\usepackage{color}
\usepackage{youngtab}
\usepackage{multirow}
\usepackage{array}
\usepackage{mathtools}
\usepackage{inputenc}
\usepackage{makecell}
\usepackage{algpseudocode}
\usepackage{tikz}
\usetikzlibrary{quantikz2}

\usepackage[colorlinks=true,linkcolor=blue,citecolor=blue,urlcolor=blue]{hyperref}

\begin{document}

\title{Floquet spintronics: tuning the current-induced spin polarization of topological surface states with light}

\author{Youngjae Kim$^{1,2}$, Aayushi Agrawal$^{2}$ and Kwon Park$^{2}$}
\affiliation{$^{1}$Department of Semiconductor Physics and Institute of Quantum Convergence Technology, Kangwon National University, Chuncheon, 24341, Korea}
\affiliation{$^{2}$School of Physics, Korea Institute for Advanced Study, Seoul 02455, Korea}

\date{\today}

\begin{abstract}

Topological surface states are a promising platform for spintronics due to spin-momentum locking.
Spin-momentum locking can induce a net spin polarization in topological surface states via an electric current, a phenomenon known as the Edelstein effect.
In this work, using Floquet theory, we show that the current-induced spin polarization of topological surface states can be tuned by illuminating them with light, thereby modifying their spin texture in momentum space.
Specifically, the electric spin susceptibility of topological surface states can be controlled and even reversed by varying the electric-field strength of high-frequency, circularly polarized light.
 
\end{abstract}

\maketitle

In conventional electronics, information is carried by electric charge, forming the basis of modern information technology.
More information, possibly even quantum information, can be carried by harnessing the other fundamental degree of freedom of electrons: spin.
Spintronics is the branch of physics in which spin is envisioned to be actively controlled and transported~\cite{Zutic04,Sinova04,Liu12,Sanchez16,Manipatruni18,Dieny20,LGuo24,YGuo24,Liu25}.

Generating spin current in devices is a central issue in spintronics.
A possible solution to this issue is to inject electrons through a ferromagnetic layer, thereby aligning spins via exchange interaction~\cite{Sarkar25,Jia25,Xie25}.
While this solution is conceptually clear, it would be preferable if spins could be manipulated electrically rather than magnetically.

Spins can be electrically manipulated in the presence of spin-orbit coupling; electric fields can alter the orbital degree of freedom, thereby modifying spin states.
Topological surface states in three-dimensional strong topological insulators are a promising platform for spintronics due to spin-momentum locking, a special realization of spin-orbit coupling~\cite{Hsieh09,Li14,Tai24}. 
A net spin polarization can be induced in spin-orbit-coupled materials, including topological surface states, by an electric current.
This current-induced spin polarization is known as the Edelstein effect~\cite{Edelstein90,Mellnik14,Johansson21,Johansson24,Hu25}.

There is an additional reason why topological surface states can serve as a particularly promising platform for spintronic devices: their robustness, stemming from topological protection.
Topological protection, however, is a double-edged sword.
On the one hand, topological surface states are robust against various local perturbations.
On the other hand, this robustness makes tuning their properties difficult.

In this work, we propose a method to overcome this difficulty by illuminating topological surface states with high-frequency, circularly polarized light.
This method is inspired by the fact that graphene, initially topologically trivial, can be transformed into a Chern insulator under illumination by high-frequency, circularly polarized light~\cite{Oka09,Mikami16,Kim19,McIver20,Merboldt25}.
Moreover, the Floquet Chern insulator emerging from illuminated graphene can change its Chern number as a function of the electric-field strength of the circularly polarized light.

\begin{figure}[]
\includegraphics[width=\columnwidth]{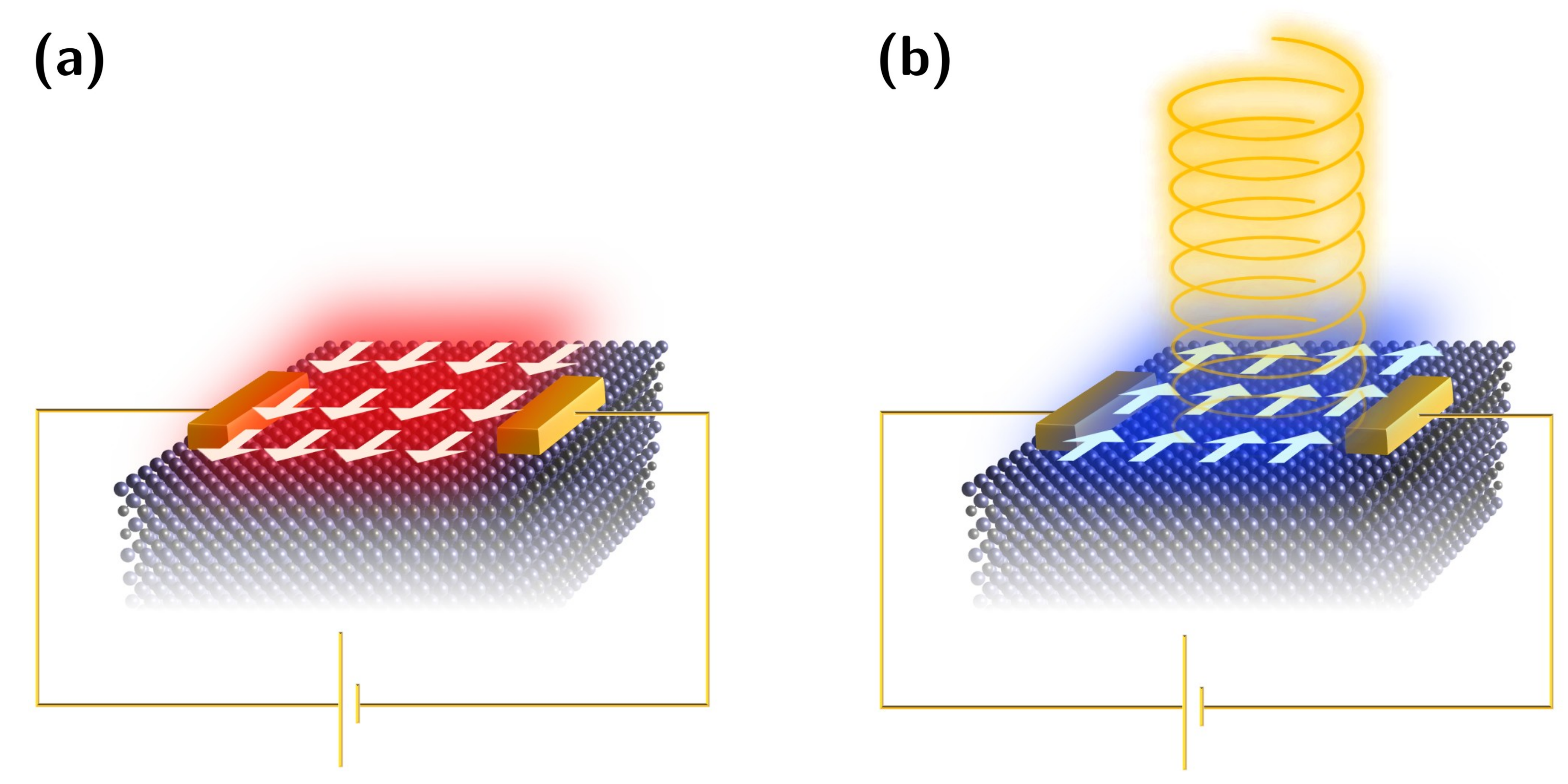}
\caption{\textbf{Schematic illustration for Floquet spintronics using topological surface states.}
({\bf a}) A net spin polarization can be induced in topological surface states by an electric current, a phenomenon known as the Edelstein effect.
({\bf b}) The electric spin polarization can be controlled and even reversed at sufficiently strong electric-field strength of high-frequency, circularly polarized light, which would open a new frontier in Floquet engineering, namely Floquet spintronics.
}
\label{FIG1}
\end{figure}

Given the similarity between topological surface states and graphene, it may be conjectured that the topological properties of topological surface states can be controlled by illuminating them with high-frequency, circularly polarized light.
In this work, using Floquet theory, we show that this is indeed true and, as an important byproduct, that the current-induced spin polarization can be tuned by varying the electric-field strength of the circularly polarized light.
See Fig.~\ref{FIG1} for an illustration.
This would open a new frontier in Floquet engineering~\cite{Oka19, Rudner20, Shan21, Sandholzer22, Castro22, Zhou23, Kobayashi23, Takahashi25}, which we call Floquet spintronics.

{\noindent {\bf Results}}

{\noindent {\bf Floquet theory of illuminated topological surface states.}} 
We begin with the effective Hamiltonian describing the topological surface states of a three-dimensional strong topological insulator, bismuth selenide (Bi$_2$Se$_3$):
$H({\bf k})=h_0({\bf k})\mathbb{I}+\sum_{\alpha=x,y,z} h_\alpha({\bf k})\sigma_\alpha$, where $\mathbb{I}$ is the identity matrix, $\sigma_\alpha$ are the Pauli matrices, and
\begin{align}
h_0({\bf k})&= \zeta+\lambda k^2,
\label{eq:h_0} \\
h_x({\bf k})&= \xi k_y (1-k^2/8),
\label{eq:h_x} \\
h_y({\bf k})&= -\xi k_x (1-k^2/8),
\label{eq:h_y} \\
h_z({\bf k})&= -\eta k_x (k_x^2-3k_y^2),
\label{eq:h_z}
\end{align}
where $k^2=k_x^2+k_y^2$ and $\zeta$, $\lambda$, $\xi$, and $\eta$ are phenomenological parameters~\cite{Baykusheva21}.
For convenience, all momenta are expressed in units of $1/a$, with $a=4.14 \textrm{\AA}$ the lattice constant. 
In this momentum unit, the numerical values of the phenomenological parameters are as follows:
(i) $\zeta= -0.0546\;\textrm{eV}$, (ii) $\lambda= 0.0779\;\textrm{eV}$, (iii) $\xi= 0.1653\;\textrm{eV}$, and (iv) $\eta= 0.0560\;\textrm{eV}$.

The time-dependent Hamiltonian describing the illuminated topological surface states under circularly polarized light can be obtained using Peierls's substitution, ${\bf k} \rightarrow {\bf k}(t)={\bf k}-{\bf A}(t)$ with ${\bf A}(t)=A(\cos{\Omega t},\sin{\Omega t})$, where $A=e E_0 a/\hbar \Omega$ with $E_0$ the electric-field strength of the light, $a$ the lattice constant, and $\Omega$ the light frequency:
\begin{align}
H({\bf k}(t))= h_0({\bf k}-{\bf A}(t))\mathbb{I}+\sum_{\alpha=x,y,z} h_\alpha({\bf k}-{\bf A}(t)) \sigma_\alpha,
\label{eq:H_time-dependent}
\end{align}
where, for convenience, we set the speed of light $c$, the electron charge $e$, and the reduced Planck constant $\hbar$ to unity from now on unless otherwise stated.

The time-dependent Hamiltonian $H({\bf k}(t))$ can be analyzed using Floquet theory. 
Specifically, solving the time-dependent Schr\"{o}dinger equation amounts to diagonalizing the Floquet Hamiltonian:
\begin{align}
[H_{\rm F}({\bf k})]_{nm}=H_{nm}({\bf k})-n\Omega\delta_{nm},
\label{eq:Floquet_Hamiltonian}
\end{align}
where $H_{nm}({\bf k})=\int_0^T \frac{dt}{T} H({\bf k}(t)) e^{i(n-m)\Omega t}$ with $T=2\pi/\Omega$.
Mathematically, the Floquet Hamiltonian can be diagonalized by a unitary matrix ${\cal U}_{\bf k}$: 
\begin{equation}
{\cal U}^{\dagger}_{\bf k} H_{\rm F}({\bf k}){\cal U}_{\bf k} = {\rm diag}[\epsilon_{\mu n}({\bf k})],
\label{eq:U_k}
\end{equation}
where $\epsilon_{\mu n}({\bf k})$ denotes the quasienergy of the Floquet eigenstates, i.e., $\epsilon_{\mu n}({\bf k})=\epsilon_{\mu}({\bf k})+n\Omega$, with $\mu=\pm$ labeling the upper/lower quasienergy subbands and $n$ labeling the Floquet index.
The quasienergy $\epsilon_{\mu n}({\bf k})$ and the unitary matrix ${\cal U}_{\bf k}$ can be used to construct the nonequilibrium Green's functions, which, in turn, determine the electric response of spin polarization in illuminated topological surface states.

The nonequilibrium Green's functions can be constructed step by step.
First, the noninteracting Green's functions can be obtained as follows:
\begin{align}
G^{r,<}_{0}({\bf k},\omega) = {\cal U}_{\bf k} {\cal G}_{\rm F}^{r,<}({\bf k},\omega) {\cal U}^{\dagger}_{\bf k} ,
\label{eq:G_0}
\end{align}
where $G^{r,<}_0(\mathbf{k},\omega)$ denote the noninteracting retarded and lesser Green's functions for the Bloch state, respectively, and ${\cal G}^{r,<}_{\rm F}(\mathbf{k},\omega)$ denote the corresponding Green's functions for the Floquet eigenstate. 
Specifically, the retarded Green's function for the Floquet eigenstate is given by:
\begin{align}
[{\cal G}_{\rm F}^r({\bf k},\omega)]_{\mu n,\nu m}= \frac{1}{\omega-\mu_0-\epsilon_{\mu}({\bf k})-n\Omega+i\delta}\delta_{\mu\nu}\delta_{nm} ,
\label{eq:G_F^r}
\end{align}
where $\omega$ is restricted within the first Floquet Brillouin zone, i.e., $-\Omega/2 < \omega \le \Omega/2$, and $\mu_0$ is the chemical potential.
In this work, $\mu_0$ is set to the Fermi level typically observed in experiments, so the Fermi surface lies well within the conduction band.
Specifically, $\mu_0$ is set to 0.28 eV above the Dirac point in the absence of illumination, i.e., at $A=0$ \cite{Bianchi10}.
Under illumination, $\mu_0$ is adjusted to maintain the same electron density.

The lesser Green's function for the Floquet eigenstate can be obtained using the Floquet eigenstate-wise thermalization (FEWT) hypothesis~\cite{Kim25}:
\begin{align}
{\cal G}_{\rm F}^<(\mathbf{k},\omega)= \left( {\cal G}_{\rm F}^{a}(\mathbf{k},\omega)-{\cal G}_{\rm F}^{r}(\mathbf{k},\omega) \right) \mathcal{F}_{\rm FD}(\mathbf{k},\omega) ,
\label{eq:G_F^<}
\end{align}
where ${\cal G}_{\rm F}^{a}({\bf k},\omega)=[{\cal G}_{\rm F}^{r}({\bf k},\omega)]^\dagger$ is the advanced Green's function for the Floquet eigenstate, and
\begin{align}
[{\cal F}_{\rm FD}({\bf k},\omega)]_{\mu n,\nu m}=f_{\rm FD} \left(
\omega-\mu_0-\langle {\cal N} \rangle_{\mu n {\bf k}} \Omega \right)
\delta_{\mu\nu}\delta_{nm} ,
\label{eq:cal_F}
\end{align}
where $f_{\rm FD}(\epsilon)$ is the usual Fermi-Dirac distribution function with the temperature set to zero in this work. 
In the FEWT hypothesis, the chemical potential of each Floquet eigenstate is shifted from $\mu_0$ by the expectation value of the Floquet index operator: $\langle {\cal N} \rangle_{\mu n {\bf k}}=\langle \phi_{\mu n} ({\bf k}) | {\cal N} | \phi_{\mu n} ({\bf k}) \rangle$, where $[{\cal N}]_{\mu n,\nu m}=n\delta_{\mu\nu}\delta_{nm}$ is the Floquet index operator, and $|\phi_{\mu n}({\bf k})\rangle$ is the Floquet eigenstate with quasienergy eigenvalue $\epsilon_{\mu n}({\bf k})$.
In the high-frequency limit, where quasienergy bands are well separated, $\langle {\cal N} \rangle_{\mu n {\bf k}}$ reduces to $n$, thereby reproducing the thermalization scheme used for the microwave-irradiated quantum Hall states~\cite{Shi03, Durst03, Park04}. 
In this work, we set $\Omega$ to 8 eV, which is 1,000 times the minimum energy gap of the quasienergy bands, as shown later.

Next, we compute the fully interacting Green's functions, which account for impurity scattering, using the Keldysh-Dyson equation:
\begin{align}
G^r &= \left( [G_0^r]^{-1} - \Sigma^r \right)^{-1} , \\
G^< &= G^{r} \left( [G_0^r]^{-1} G_0^< [G_0^a]^{-1} + \Sigma^{<} \right) G^{a} ,
\label{eq:Keldysh-Dyson}
\end{align}
where $G^a_0=[G^r_0]^\dagger$ and $G^a=[G^r]^\dagger$ denote the noninteracting and fully interacting advanced Green's functions for the Bloch state, respectively, and $\Sigma^{r,<}$ represents the retarded and lesser self-energies.
Note that the arguments of the Green's functions, ${\bf k}$ and $\omega$, are omitted for brevity.
In this work, the self-energies are calculated using the self-consistent Born approximation (SCBA) of delta-function impurity scattering~\cite{Lee14,Hwang21}: 
\begin{equation}
[\Sigma^{r,<}(\omega)]_{an,bm} = v_\mathrm{imp}^2 \int \frac{d^{2}{\mathbf{k}}}{(2\pi)^2} [G^{r,<}(\mathbf{k},\omega)]_{an,am} \delta_{ab} ,
\label{eq:SCBA}
\end{equation}
where $a$ and $b$ denote spins (i.e., $\uparrow$ or $\downarrow$) and $v_\mathrm{imp}$ represents the strength of impurity scattering. 
Here, we set $v_\mathrm{imp}=1.0$ unless stated otherwise.
Note that self-energies are diagonal in the spin index because impurity scattering is assumed to be spin-independent.

Once the nonequilibrium Green's functions are obtained as a solution of the Keldysh-Dyson equation, the electric response of spin polarization can be computed within linear response theory:
\begin{equation}
\langle S_\alpha \rangle = \sum_\beta \chi^{\rm s}_{\alpha\beta} E_\beta,
\label{eq:electric_spin_response}
\end{equation}
where $S_\alpha$ is the spin density operator along the $\alpha$-direction, $E_\beta$ is the external electric field along the $\beta$-direction, and $\chi^{\rm s}_{\alpha\beta}$ is the electric spin susceptibility relating the two.
The electric spin susceptibility can be computed using the Kubo formula. 
Ignoring the vertex correction, the Kubo formula for electric spin susceptibility can be written as follows:
\begin{align}
\chi^{\rm s}_{\alpha\beta} =&\int \frac{d^2 {\bf k}}{(2\pi)^2} \int \frac{d\omega}{2\pi}
{\rm Tr} \Big( S_\alpha \frac{\partial G^r({\bf k},\omega)}{\partial \omega} J_\beta({\bf k}) G^<({\bf k},\omega)
\nonumber \\
&-S_\alpha G^<({\bf k},\omega)J_\beta({\bf k})\frac{\partial G^a({\bf k},\omega)}{\partial\omega} \Big),
\label{eq:Kubo}
\end{align}
where $J_\beta({\bf k})$ is the current density operator along the $\beta$-direction in the Floquet representation: $[J_\beta({\bf k})]_{nm}=\int_0^T \frac{dt}{T} \frac{\partial H({\bf k}(t))}{\partial k_\beta} e^{i(n-m)\Omega t}$.
Note that Equation~\eqref{eq:Kubo} is a Floquet generalization of the equilibrium expression~\cite{Liu08}, which can be derived following the derivation of the similar nonequilibrium Kubo formula~\cite{Kim25}.

\begin{figure}[]
\includegraphics[width=0.9\columnwidth]{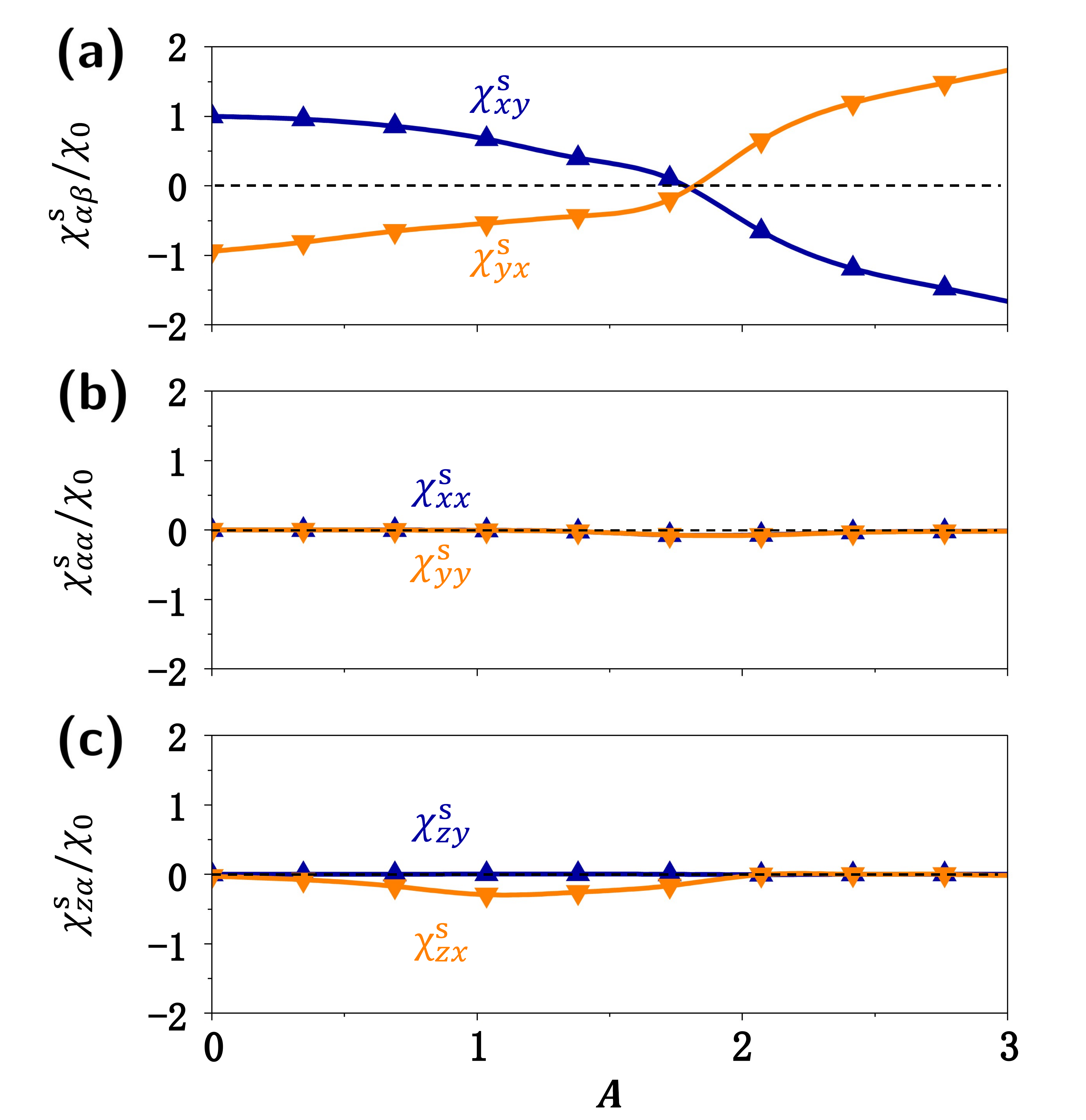}
\caption{\textbf{Electric spin susceptibility.} 
The electric spin susceptibility is plotted as a function of the normalized electric-field strength $A$ of high-frequency, circularly polarized light.
Here, $A=e E_0 a /\hbar \Omega$, where $e$ is the electron charge, $E_0$ is the electric-field strength, $a$ is the lattice constant, and $\Omega$ is the light frequency.
Throughout this work, the light frequency $\Omega$ is set to 8 eV, which is 1,000 times the minimum energy gap of the quasienergy bands.
({\bf a}) Transverse electric spin susceptibilities, $\chi^{\rm s}_{xy}$ and $\chi^{\rm s}_{yx}$, as a function of $A$.
({\bf b}) Longitudinal electric spin susceptibilities, $\chi^{\rm s}_{xx}$ and $\chi^{\rm s}_{yy}$, as a function of $A$.
({\bf c}) Out-of-plane electric spin susceptibilities, $\chi^{\rm s}_{zx}$ and $\chi^{\rm s}_{zy}$, as a function of $A$.
All electric spin susceptibilities are scaled by $\chi_0$, the zero-field transverse electric spin susceptibility in the absence of illumination, i.e., $\chi_0=|\chi^{\rm s}_{\alpha\beta}(A=0)|$.
In physical units, $\chi_0$ is $8.97\times10^{-10}$ $\mu_B\cdot{\rm m/V}$ when defined as the induced magnetization per unit cell in response to an electric field, rather than the induced spin per unit cell.
This value is roughly on the same order of magnitude as those computed for other materials~\cite{Johansson21}.
}
\label{FIG2}
\end{figure}

Figure~\ref{FIG2} shows the electric spin susceptibility as a function of the normalized electric-field strength $A$ of high-frequency, circularly polarized light.  
As shown in Fig.~\ref{FIG2} ({\bf a}), the transverse electric spin susceptibilities, $\chi^{\rm s}_{xy}$ and $\chi^{\rm s}_{yx}$, change sign at $A \simeq 1.8$.
This indicates that the current-induced spin polarization can be controlled and even reversed by varying the electric-field strength of high-frequency, circularly polarized light, as noted at the beginning of this paper.
Meanwhile, the longitudinal electric spin susceptibilities, $\chi^{\rm s}_{xx}$ and $\chi^{\rm s}_{yy}$, are essentially zero regardless of $A$.
There is a slight response in the out-of-plane electric spin susceptibility, $\chi^{\rm s}_{zx}$, whereas the other out-of-plane electric spin susceptibility, $\chi^{\rm s}_{zy}$, is essentially zero.

Now, a natural question arises: why does the transverse electric spin susceptibility change sign as a function of the electric-field strength of high-frequency, circularly polarized light?
To answer this question, we need to analyze the Floquet Hamiltonian more closely in the high-frequency limit.

\begin{figure*}[]
\includegraphics[width=0.9\textwidth]{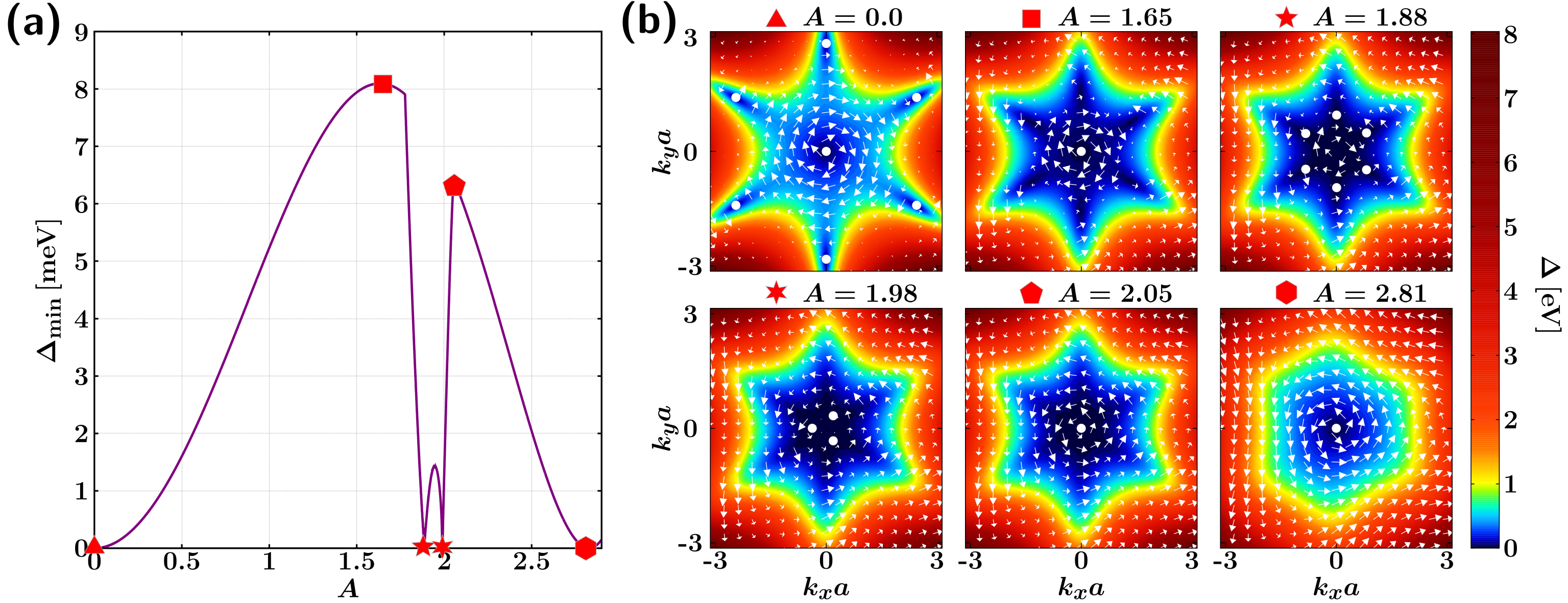}
\caption{\textbf{Minimum energy gap and evolution of spin texture.}
({\bf a}) Minimum energy gap $\Delta_{\rm min}$ as a function of the normalized electric-field strength $A$ of high-frequency, circularly polarized light.
({\bf b}) In-plane components of the spin texture in momentum space, ${\bf s}_\parallel({\bf k})$, for the conduction band at different values of $A$, indicated by the following symbols: (i) triangle, $A=0$; (ii) square, $A=1.65$; (iii) star, $A=1.88$; (iv) David's star, $A=1.98$; (v) pentagon, $A=2.05$; and (vi) hexagon, $A=2.81$.
Note that the positions of the minimum energy gap are indicated by solid white circles.
}
\label{FIG3}
\end{figure*}

\begin{figure*}[]
\includegraphics[width=0.9\textwidth]{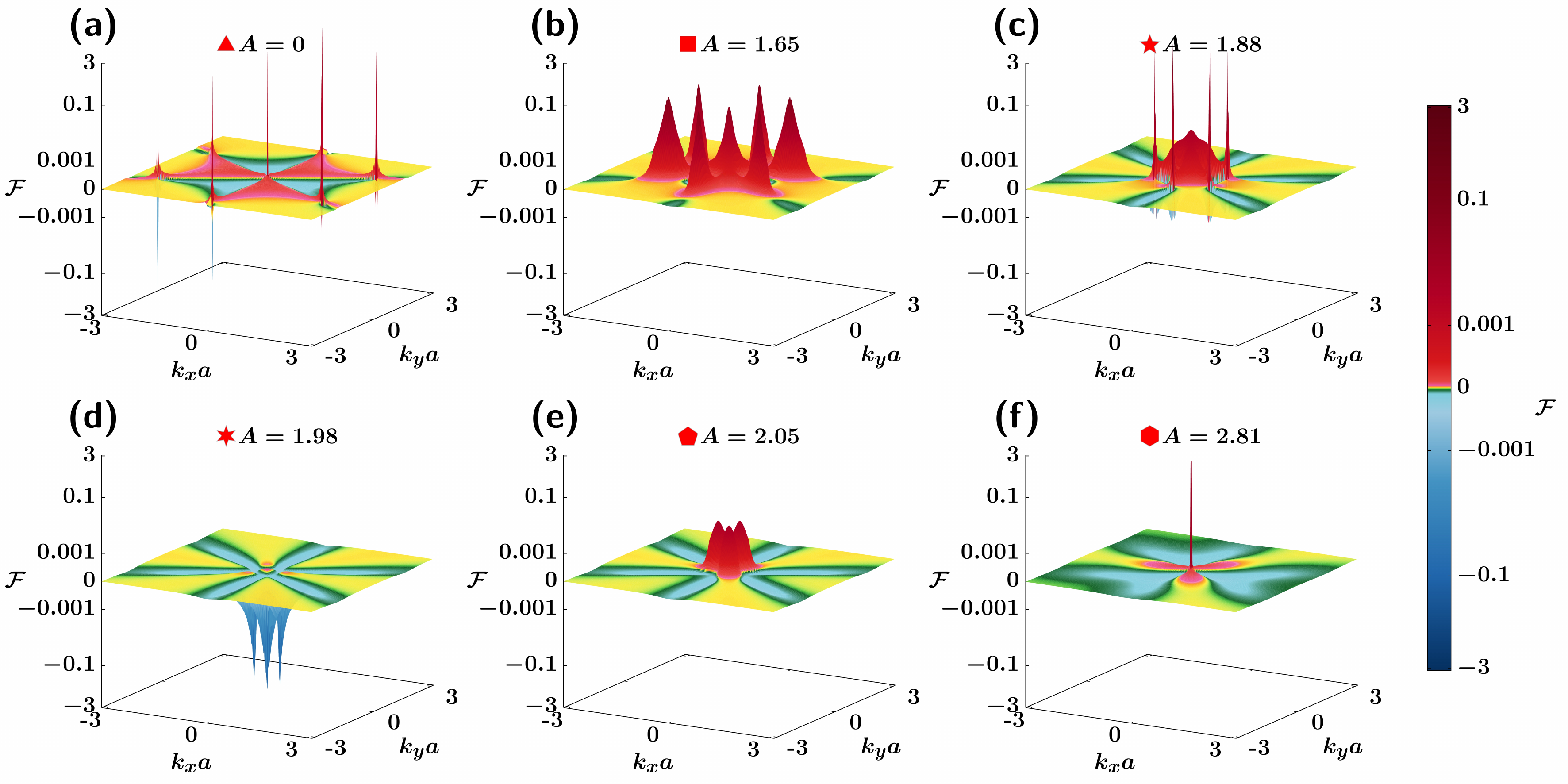}
\caption{\textbf{Berry curvature and topological phase transitions.} 
The Berry curvature ${\cal F}$ is plotted as a function of the normalized electric-field strength $A$ of high-frequency, circularly polarized light.
({\bf a}) The Berry curvature exhibits sharp peaks at the positions of the minimum energy gap in the absence of illumination, i.e., at $A=0$.
({\bf b}) The Berry curvature becomes predominantly positive at $A=1.65$.
({\bf c}) The Berry curvature remains overall positive, but negative regions grow at $A=1.88$.
({\bf d}) The Berry curvature becomes predominantly negative at $A=1.98$.
({\bf e}) The Berry curvature forms broad peaks around the $\Gamma$ point, which are surrounded by sizable negative sixfold channels at $A=2.05$.
({\bf f}) The Berry curvature becomes small except for a sharp peak at the $\Gamma$ point at $A=2.81$.
}
\label{FIG4}
\end{figure*}

{\noindent {\bf High-frequency effective Hamiltonian.}} 
At high frequency, the Floquet Hamiltonian greatly simplifies: it reduces to a set of well-separated, independent effective Hamiltonians.
\begin{align}
[H_{\rm F}({\bf k})]_{nm}=H_{\rm high}({\bf k})-n\Omega\delta_{nm},
\label{eq:Floquet_high}
\end{align}
where, according to the Floquet-Magnus expansion~\cite{Bukov15,Eckardt15},
\begin{align}
H_{\rm high}({\bf k})=\tilde{H}_0({\bf k})+\sum_{m=1}^{\infty} \frac{1}{m\Omega}[\tilde{H}_{-m}({\bf k}),\tilde{H}_m({\bf k})],
\label{eq:H_high}
\end{align}
with $\tilde{H}_{m}({\bf k})=\int_0^T \frac{dt}{T} H({\bf k}(t)) e^{-im\Omega t}$.

After some algebra, as explained in Appendix~\ref{appen:H_high}, the high-frequency effective Hamiltonian can be expressed in the following matrix form:
\begin{align}
H_{\rm high}({\bf k}) 
&= 
\left(
\begin{array}{cc}
h_{00}({\bf k})+h_{z}({\bf k}) & h_{-}({\bf k})  \\
h_{+}({\bf k}) & h_{00}({\bf k})-h_{z}({\bf k})
\end{array}
\right)
\nonumber \\
&+\frac{1}{\Omega}
\left(
\begin{array}{cc}
\Pi_{\bf k} & \Delta_{\bf k} \\
\Delta^*_{\bf k} & -\Pi_{\bf k} 
\end{array}
\right),
\label{eq:H_high_mat}
\end{align}
where $h_z({\bf k})$ is defined in Eq.~\eqref{eq:h_z}, and 
\begin{align}
h_{00}({\bf k})&=\zeta+\lambda(k^2+A^2), 
\label{eq:h_00} \\
h_{\pm}({\bf k})&=\mp i\xi k_{\pm}(\Gamma-A^2/8),
\label{eq:h_pm}
\end{align}
and 
\begin{align}
\Pi_{\bf k}&=\xi^2 A^2 \left( \Gamma\left(\Gamma-\frac{1}{4}k^2\right) +\frac{1}{128}A^2 k^2 \right), 
\label{eq:Pi} \\
\Delta_{\bf k}&=i\frac{3}{8}\xi\eta A^2 \left( k_{-}^4 +8k_{+}^2\left(\Gamma-\frac{1}{8}k^2-\frac{1}{16}A^2\right) \right),
\label{eq:Delta}
\end{align}
with $k_{\pm}=k_x \pm i k_y$ and $\Gamma=1-(k^2+A^2)/8$.

The high-frequency quasienergy bands $\epsilon_{\pm}({\bf k})$ can be obtained by diagonalizing $H_{\rm high}({\bf k})$.
Because the quasienergies are well separated at high frequency, the high-frequency quasienergy bands can be treated as the usual equilibrium energy bands, which are filled up to the chemical potential $\mu_0$.
For the value of $\mu_0$ mentioned earlier, the high-frequency conduction band $\epsilon_{+}({\bf k})$ is filled near the $\Gamma$ point, and its spin texture determines the electric spin susceptibility of the illuminated topological surface states.

{\noindent {\bf Spin texture reversal and topological phase transitions.}} 
The spin texture in momentum space, i.e., the spin expectation value of the Bloch states as a function of ${\bf k}$, can be obtained from the high-frequency effective Hamiltonian, which can be expressed in the following simplified notation:
\begin{align}
H_{\rm high}({\bf k}) &=
\tilde{h}_{0}({\bf k})\mathbb{I}+\sum_{\alpha=x,y,z} \tilde{h}_{{\alpha}}({\bf k}) \sigma_{\alpha},
\label{eq:H_high_simple}
\end{align}
where the explicit forms of $\tilde{h}_{0}({\bf k})$ and $\tilde{h}_{\alpha}({\bf k})$ can be derived from Eq.~\eqref{eq:H_high_mat}.

Due to the Zeeman-coupling structure of the high-frequency effective Hamiltonian, the spin texture of the high-frequency conduction band is given by 
${\bf s}({\bf k})\equiv \langle\phi_{+}({\bf k})|\boldsymbol{\sigma}|\phi_{+}({\bf k})\rangle = \tilde{\bf h}({\bf k})/|\tilde{\bf h}({\bf k})|$, where 
$\phi_{+}({\bf k})$ is the Bloch eigenstate of the conduction band at ${\bf k}$, and $\tilde{\bf h}({\bf k})=(\tilde{h}_x({\bf k}),\tilde{h}_y({\bf k}),\tilde{h}_z({\bf k}))$.
In this work, we are mainly interested in the in-plane components of the spin texture, ${\bf s}_\parallel({\bf k})=(\tilde{h}_x({\bf k}),\tilde{h}_y({\bf k}))/|\tilde{\bf h}({\bf k})|$, because the out-of-plane components contribute little to the electric spin susceptibility. 
The change in sign of the electric spin susceptibility indicates that the in-plane components of the spin texture are modified by varying the electric-field strength of the circularly polarized light.
The next step is to investigate whether this is indeed the case.

Figure~\ref{FIG3} shows the minimum energy gap and the evolution of the spin texture for the conduction band as a function of $A$.
For $A \lesssim 1.65$, the in-plane components of the spin texture rotate overall clockwise, and the minimum energy gap, occurring at the $\Gamma$ point, increases smoothly with $A$. 
Beyond that, the energy gap collapses sharply across two consecutive transitions at $A=1.88$ (sixfold Dirac-like band touching) and $A=1.98$ (threefold Dirac-like band touching) as the in-plane components of the spin texture begin rotating counterclockwise. 
For $2.05 \lesssim A \lesssim 2.81$, the in-plane components of the spin texture rotate overall counterclockwise, and the minimum energy gap, again occurring at the $\Gamma$ point, decreases smoothly with $A$. 
Consequently, we conclude that this reversal of spin texture causes the transverse electric spin susceptibilities, $\chi^{s}_{xy}$ and $\chi^{\rm s}_{yx}$, to change sign.

Why does this reversal of spin texture occur?
As mentioned at the beginning of this paper, it is conjectured that circularly polarized light can induce topological phase transitions in topological surface states under sufficiently strong electric fields, thereby altering the spin texture.
To confirm this, we investigate how the Berry curvature changes as a function of $A$.
Note that, for a given spin texture, the Berry curvature can be computed as ${\cal F}({\bf k})={\bf s}({\bf k}) \cdot\partial {\bf s}({\bf k}) / \partial k_x \times \partial {\bf s}({\bf k}) /\partial k_y$.

Figure~\ref{FIG4} shows the Berry curvature of the conduction band as a function of $A$.
The most important point to note is that the overall behavior of the Berry curvature changes dramatically as it passes through $A=1.88$ and $A=1.98$.
Below $A=1.65$, the Berry curvature is predominantly positive.
At the critical points $A=1.88$ and $A=1.98$, where the quasienergy gap closes through two Dirac-like band touchings, the Berry curvature changes from predominantly positive to negative. 
Above $A=2.05$, the Berry curvature has roughly equal regions of positive and negative values.
This indicates that the two Dirac-like band touchings induce topological phase transitions, which consequently reverse the spin texture.

{\noindent {\bf Conclusion}} 

In this work, we show that the current-induced spin polarization can be tuned by illuminating topological surface states with light.
Specifically, the electric spin susceptibility of topological surface states can be controlled and even reversed by varying the electric-field strength of high-frequency, circularly polarized light.
This phenomenon results from the reversal of spin texture in momentum space, which is induced by topological phase transitions. 
Tuning the current-induced spin polarization with light can open a new frontier in Floquet engineering, which we call Floquet spintronics.

For experimental realization, let us estimate the parameter ranges.
First, the high-frequency limit requires that $\Omega$ be sufficiently large.
We take $\Omega$ to be 8 eV (roughly 2,000 THz) as a conservative estimate, but it can be set lower, say 4 eV or 1,000 THz, which can be generated by an ultraviolet (UV-B) laser.
Second, the normalized electric-field strength $A=e E_0 a/\hbar\Omega$ should be sufficiently large to reverse the spin texture.
Nowadays, the electric field of the light can be applied at the order of 0.1 V/\AA~\cite{Schiffrin13,Mayer15,Higuchi17,Uchida22,Merboldt25}.
Given that the lattice spacing of topological surface states is 4.14 \AA, $A$ can reach the order of unity in the near future if the electric field is increased to the order of 1 V/\AA.

\noindent {\bf Acknowledgments}

The authors thank the Center for Advanced Computation (CAC) at Korea Institute for Advanced Study (KIAS) for providing computing resources for this work.
This work is partially supported by the KIAS Individual Grants, PG098801 (A.A.) and PG032303 (K.P.). 
Additionally, Y.K. is supported by 2026 Research Grant from Kangwon National University and by a grant from the National Research Foundation of Korea (NRF) funded by the Ministry of Science and ICT of the Korean government (RS-2025-00553820).


\appendix

\section{Derivation of the high-frequency effective Hamiltonian}
\label{appen:H_high}

We begin by rewriting the time-dependent Hamiltonian describing the illuminated topological surface states under circularly polarized light as follows:
\begin{align}
H({\bf k}(t))= h_0({\bf k}-{\bf A}(t))\mathbb{I}+\sum_{\alpha=x,y,z} h_\alpha({\bf k}-{\bf A}(t)) \sigma_\alpha,
\label{eq:appen_H_time-dependent}
\end{align}
where
\begin{align}
h_x({\bf k}-{\bf A}(t))&= \xi (k_y-A\sin{\Omega t}) 
\nonumber \\
&\times\left[\Gamma+\frac{A}{4}(k_x \cos{\Omega t} +k_y \sin{\Omega t}) \right],
\label{eq:h_x_time-dependent} \\
h_y({\bf k}-{\bf A}(t))&= -\xi (k_x-A\cos{\Omega t}) 
\nonumber \\
&\times\left[\Gamma+\frac{A}{4}(k_x \cos{\Omega t} +k_y \sin{\Omega t}) \right],
\label{eq:h_y_time-dependent} \\
h_z({\bf k}-{\bf A}(t))&= -\eta (k_x-A\cos{\Omega t}) 
\nonumber \\
&\times \Big[ \Lambda -2A(k_x\cos{\Omega t}-3k_y\sin{\Omega t})
\nonumber \\ 
&+A^2(\cos^2{\Omega t}-3\sin^2{\Omega t})\Big],
\label{eq:h_z_time-dependent}
\end{align}
with $\Gamma=1-(k^2+A^2)/8$ and $\Lambda=k^2_x-3k^2_y$.

As noted in the main text, the high-frequency effective Hamiltonian can be derived via the Floquet-Magnus expansion:
\begin{align}
H_{\rm high}({\bf k})&= H_{\rm high}^{(0)}({\bf k})+H_{\rm high}^{(1)}({\bf k})
\nonumber \\
&=\tilde{H}_0({\bf k})+\sum_{m=1}^{\infty} \frac{1}{m\Omega}[\tilde{H}_{-m}({\bf k}),\tilde{H}_m({\bf k})],
\label{eq:appen_H_high}
\end{align}
with $\tilde{H}_{m}({\bf k})=\int_0^T \frac{dt}{T} H({\bf k}(t)) e^{-im\Omega t}$.

Specifically, the zeroth-order high-frequency effective Hamiltonian in $1/\Omega$ is given as
\begin{align}
H_{\rm high}^{(0)}({\bf k})=\tilde{H}_0({\bf k})=h_{0 0}({\bf k})\mathbb{I}+\sum_{\alpha=x,y,z} h_{\alpha 0}({\bf k}) \sigma_\alpha,
\label{eq:H0_high}
\end{align}
where $h_{\alpha 0}({\bf k})=\int_0^T \frac{dt}{T} h_\alpha({\bf k}-{\bf A}(t))$; their explicit expressions are as follows:
\begin{align}
h_{00}({\bf k})&=\zeta+\lambda(k^2+A^2), 
\label{eq:appen_h_00} \\
h_{x0}({\bf k})&=\xi k_y(\Gamma-A^2/8), 
\label{eq:appen_h_x0} \\
h_{y0}({\bf k})&=-\xi k_x(\Gamma-A^2/8), 
\label{eq:appen_h_y0} \\
h_{z0}({\bf k})&=-\eta k_x \Lambda=h_{z}({\bf k}).
\label{eq:appen_h_z0}
\end{align}
Consequently, $H_{\rm high}^{(0)}({\bf k})$ can be written in the following matrix form:
\begin{align}
H_{\rm high}^{(0)}({\bf k})
&=
\left(
\begin{array}{cc}
h_{00}({\bf k})+h_{z}({\bf k}) & h_{-}({\bf k}) \\
h_{+}({\bf k}) & h_{00}({\bf k})-h_{z}({\bf k}) 
\end{array}
\right),
\label{eq:H0_high_mat}
\end{align}
where $h_{\pm}({\bf k})=h_{x 0}({\bf k}) \pm i h_{y 0}({\bf k})=\mp i\xi k_{\pm}(\Gamma-A^2/8)$ and $k_{\pm}=k_x \pm i k_y$.

Meanwhile, the first-order high-frequency effective Hamiltonian in $1/\Omega$ is given as
\begin{align}
H_{\rm high}^{(1)}({\bf k})&=\sum_{m=1}^{\infty} \frac{1}{m\Omega}[\tilde{H}_{-m}({\bf k}),\tilde{H}_m({\bf k})]
\nonumber \\
&=2i \sum_{m=1}^\infty \frac{1}{m\Omega} \sum_{\alpha,\beta,\gamma}  \epsilon_{\alpha\beta\gamma} h_{\alpha, -m}({\bf k}) h_{\beta m}({\bf k})  \sigma_\gamma,
\label{eq:H1_high}
\end{align}
where $\epsilon_{\alpha\beta\gamma}$ is the Levi-Civita symbol and $h_{\alpha m}({\bf k})=\int_0^T \frac{dt}{T} h_\alpha({\bf k}-{\bf A}(t))e^{-im\Omega t}$.
Note that due to the structure of $h_\alpha({\bf k}-{\bf A}(t))$, it suffices to sum over $m$ only up to $m=2$.
Explicit expressions for $h_{\alpha m}({\bf k})$ are given as follows:
\begin{align}
h_{x 1}({\bf k})&=\xi \frac{A}{8} \Big(k_x k_y +i(4\Gamma-k_y^2)\Big), 
\label{eq:h_x1} \\
h_{y 1}({\bf k})&=i\xi \frac{A}{8} \Big(k_x k_y -i(4\Gamma-k_x^2)\Big), 
\label{eq:h_y1} \\
h_{z 1}({\bf k})&=\frac{3}{2}\eta A k_{+}^2,
\label{eq:h_z1}
\end{align}
and
\begin{align}
h_{x 2}({\bf k})&=i\xi \frac{A^2}{16} k_{-}, 
\label{eq:h_x2} \\
h_{y 2}({\bf k})&=\xi \frac{A^2}{16} k_{-}, 
\label{eq:h_y2} \\
h_{z 2}({\bf k})&=-\frac{3}{2} \eta A^2 k_{+}.
\label{eq:h_z2}
\end{align}
Note that $h_{\alpha,-m}({\bf k})=h^*_{\alpha m}({\bf k})$.
After some algebra, $H_{\rm high}^{(1)}({\bf k})$ can be written in the following matrix form:
\begin{align}
H_{\rm high}^{(1)}({\bf k})&=
\frac{1}{\Omega}
\left(
\begin{array}{cc}
\Pi_{\bf k} & \Delta_{\bf k} \\
\Delta^*_{\bf k} & -\Pi_{\bf k} 
\end{array}
\right),
\label{eq:H1_high_mat}
\end{align}
where
\begin{align}
\Pi_{\bf k}&=\xi^2 A^2 \left( \Gamma\left(\Gamma-\frac{1}{4}k^2\right) +\frac{1}{128}A^2 k^2 \right), 
\label{eq:appen_Pi} \\
\Delta_{\bf k}&=i\frac{3}{8}\xi\eta A^2 \left( k_{-}^4 +8k_{+}^2\left(\Gamma-\frac{1}{8}k^2-\frac{1}{16}A^2\right) \right).
\label{eq:appen_Delta}
\end{align}

Finally, combining everything, the high-frequency effective Hamiltonian can be expressed in the following matrix form:
\begin{align}
H_{\rm high}({\bf k})&= 
\left(
\begin{array}{cc}
h_{00}({\bf k})+h_{z}({\bf k}) & h_{-}({\bf k})  \\
h_{+}({\bf k}) & h_{00}({\bf k})-h_{z}({\bf k})
\end{array}
\right)
\nonumber \\
&+\frac{1}{\Omega}
\left(
\begin{array}{cc}
\Pi_{\bf k} & \Delta_{\bf k} \\
\Delta^*_{\bf k} & -\Pi_{\bf k} 
\end{array}
\right).
\label{eq:appen_H_high_mat}
\end{align}
which is identical to Eq.~\eqref{eq:H_high_mat}.

\bibliography{references}

\end{document}